\documentclass{emulateapj}

\usepackage{graphicx}
\usepackage{subfigure}
\usepackage{amssymb}
\usepackage{psfrag}

\shorttitle{WMAP Non-Gaussianity Due to Possible Foreground
Signals} \shortauthors{X. Liu and S. N. Zhang}

\begin{document}


\title{Non-Gaussianity Due to Possible Residual Foreground Signals in\\
{\it Wilkinson Microwave Anisotropy Probe} First-Year Data \\
Using Spherical Wavelet Approaches}


\author{Xin Liu\altaffilmark{1} and Shuang Nan Zhang\altaffilmark{1,2,3,4}}
\altaffiltext{1}{Physics Department and Tsinghua Center for
Astrophysics, Tsinghua University, Beijing 100084, China;
liux00@mails.tsinghua.edu.cn, zhangsn@tsinghua.edu.cn.}
\altaffiltext{2}{Key Laboratory of Particle Astrophysics,
Institute of High Energy Physics, Chinese Academy of Sciences,
P.O. Box 918-3, Beijing 100039, China}
\altaffiltext{3}{Physics Department, University of Alabama in
Huntsville, Huntsville, AL 35899, USA}
\altaffiltext{4}{Space Science Laboratory, NASA Marshall Space
Flight Center, SD50, Huntsville, AL 35812, USA}



\begin{abstract}
We perform multi-scale non-Gaussianity detection and localization
to the {\it Wilkinson Microwave Anisotropy Probe} ({\it WMAP})
first-year data in both wavelet and real spaces. Such an analysis
is facilitated by spherical wavelet transform and inverse
transform techniques developed by the YAWtb team. Skewness and
kurtosis as test statistics are calculated on scales from about
$1^{\circ}$ to $30^{\circ}$ on the sky as well as toward different
directions using anisotropic spherical Morlet wavelet (SMW). A
maximum deviation from Gaussian simulations with a right tail
probability of $\sim 99.9\%$ is detected at an angular scale of
$\sim12^{\circ}$ at an azimuthal orientation of $\sim 0^{\circ}$
on the sky. In addition, some significant non-Gaussian spots have
been identified and localized in real space from both the combined
Q-V-W map recommended by the {\it WMAP} team and the Tegmark
foreground-cleaned map. Systematic effects due to beams and noise
can be rejected as the source of this non-Gaussianity. Several
tests show that residual foreground contamination may
significantly contribute to this non-Gaussian feature. It is thus
still premature to do more precise tests on the non-Gaussianity of
the intrinsic CMB fluctuations before we can identify the origin
of these foreground signals, understand their nature, and finally
remove them from the CMB maps completely.

\end{abstract}





\section{Introduction}

Since the release of the {\it Wilkinson Microwave Anisotropy
Probe} ({\it WMAP}) first-year data, much work has been done to
investigate the properties of whole-sky cosmic microwave
background (CMB) anisotropies with high resolution and sensitivity
(Bennett {\it et al.} 2003a; Hinshaw {\it et al.} 2003a,b). The
currently favored $\Lambda$CDM model is based on the assumption of
Gaussian initial fluctuations generated by inflation, which result
in Gaussian temperature anisotropies in the CMB (Guth \& Pi 1982;
Hawking 1982; Bardeen, Steinhardt \& Turner 1983). Thus the test
of non-Gaussianity of CMB temperature anisotropies provides us an
important constraint on distinguishing theories of the origin of
primordial fluctuations. In addition, the issue is also
fundamental for the determination of cosmological parameters in
the framework of the inflation paradigm, since they can only be
determined correctly from the angular power spectrum if the CMB
temperature anisotropies constitute a Gaussian random field
(Bardeen {\it et al.} 1986; Bond \& Efstathiou 1987).

Many efforts have been made to test the CMB anisotropy, and some
of them found that the WMAP data are consistent with Gaussian
primordial fluctuations
(Komatsu {\it et al.} 2003; 
Colley {\it et al.} 2003; 
Gazta\~{n}aga \& Wagg 2003; 
Gazta\~{n}aga {\it et al.} 2003), 
whereas some detected non-Gaussianity using various approaches
(Coles {\it et al.} 2004; Chiang {\it et al.} 2003; Chiang \&
Naselsky 2004; Park 2004; Eriksen {\it et al.} 2004a, 2004b;
Larson \& Wandelt 2004; Vielva {\it et al.} 2004; Hansen {\it et
al.} 2004; Mukherjee \& Wang 2004; McEwen {\it et al.} 2005; Wibig
\& Wolfendale 2005). Non-standard inflationary models and various
cosmic defects could lead to non-Gaussian primordial CMB
fluctuations. Non-Gaussianity could also be introduced by
secondary effects, such as the integrated Sachs-Wolfe effect, the
Rees-Sciama effect, the Sunyaev-Zel'dovich effect and
gravitational lensing, or by measurement systematics, foreground
and noise. Recent reports found residuals, or systematic effects
correction, as the source of non-Gaussianity (Chiang {\it et al.}
2003; Chiang \& Naselsky 2004; Naselsky {\it et al.} 2003, 2004;
Eriksen {\it et al.} 2004b; Hansen {\it et al.} 2004; Wibig \&
Wolfendale 2005). Thus, foreground removal should be dealt with
care before any conclusions on the statistical properties of the
CMB anisotropies are reached. Meanwhile, the WMAP team derived a
foreground template that is of great importance for accurately
understanding the Galactic foreground emissions.

In this paper, we perform the non-Gaussianity test on {\it WMAP}
first-year data and investigate the possible sources of the
detected non-Gaussianity, especially residual foreground
emissions, using spherical wavelet approaches. Wavelets are very
useful for data analysis due to their nature of space-frequency
localization. They have already been applied to the {\it Cosmic
Background Explorer} ({\it COBE})-Differential Microwave
Radiometer (DMR) data (Pando et al 1998; Mukherjee {\it et al.}
2000; Aghanim {\it et al.} 2001; Tenorio {\it et al.} 1999;
Barreiro {\it et al.} 2000; Cay\'{o}n {\it et al.} 2001), as well
as the {\it WMAP} first-year data (Mukherjee \& Wang, 2004; Vielva
{\it et al.} 2004; McEwen {\it et al.} 2005), using the
non-Gaussianity test in wavelet space. We perform the analysis in
both wavelet and real spaces using various wavelet bases. The
paper is organized as follows. In \S\,2, we briefly review the
general theory of spherical wavelets and its implementation. In
\S\,3, we describe how the data are processed and analyzed for the
non-Gaussianity test with spherical wavelet approaches. Results of
the test and discussions of possible sources of the detected
non-Gaussianity are presented in \S\,4. We conclude and discuss
our results further in \S\,5.


\section{Wavelets on the Sphere}

Wavelet approaches are very powerful for detecting non-Gaussianity
in CMB data (Hobson {\it et al.} 1999; Mart{\'\i}nez-Gonz{\'a}lez
{\it et al.} 2002). Due to the special nature of wavelets, a
multi-scale analysis can be done to amplify any non-Gaussian
features dominating at some specific scales. Analyzing full-sky
CMB features involves data on spherical manifolds, which require a
wavelet approach on the sphere. We consider the continuous
spherical wavelet transform (CSWT), initially proposed by Antoine
$\&$ Vandergheynst (1998), based on group theory principles, and
the framed spherical wavelet transform (FSWT), implemented by the
Yet Another Wavelet toolbox (YAWtb\footnote{Yet Another Wavelet
toolbox. See the YAWtb homepage
(http://www.fyma.ucl.ac.be/projects/yawtb) for more information.})
team. In this section we briefly review the general theory of
CSWT. We use CSWT for the non-Gaussianity detection in wavelet
space; for filtering non-Gaussian spots in real space, we use FSWT
and inverse FSWT, where the theory of frames is only useful when
the inverse transform is carried out. Since FSWT is based on CSWT
and the theory of frames is too technical to be stated in detail
here, the paper by Bogdanova I. {\it et al.} (2004) should be
referenced for details. For general readers who want to reproduce
our results, they can use \texttt{fwtsph.m} and \texttt{ifwtsph.m}
in the YAWtb for FSWT and inverse FSWT, respectively.

\subsection{The General Theory}
The CSWT, like Euclidean counterpart, is based on affine
transformations. On the two-dimensional sphere $S^2$, the basic
operations are represented by the following unitary operators
(Bogdanova {\it et al.} 2004):

$\bullet$ {\it Rigid rotation} $R_\rho$, where $\rho$ is an
element of the group of rotations $SO(3)$, may be parameterized in
terms of its Euler angles $\omega \equiv (\theta, \varphi)$ by
\begin{equation}
(R_\rho f)(\omega) = f(\rho^{-1} \omega),
\end{equation}
where $\theta \in [0,\pi]$, $\varphi \in [0,2\pi)$, $f$ is an
operator, and $\rho^{-1}$ is the opposite operator of $\rho$ in
the group of rotations $SO(3)$.

$\bullet$ {\it Conformal dilation} $D_a$, with scale $a \in
\Re^{+}$:
\begin{equation}
(D_a f)(\omega) = \lambda(a,\theta)^{1/2} f(\omega_{1/a}),
\end{equation}
where $\Re^{+}$ is the set of positive real numbers, $\omega_a
\equiv (\theta_a, \varphi)$, and the corresponding dilation size
on the sky is given by
\begin{equation}\label{size}
\tan\frac{\theta_a}{2} = a\tan\frac{\theta}{2},
\end{equation}
where $\lambda$ is a normalization factor which is given by
\begin{equation}
\lambda(a,\theta) = \frac{4 a^2}
{[(a^2-1)\cos{\theta}+(a^2+1)]^2}.
\end{equation}
A set of wavelet basis on the sphere is constructed by rotations
and
dilations %
of an admissible mother spherical wavelet $\psi \in L^2(S^2)$. The
corresponding wavelet family $\{\psi_{a,\rho} \equiv R_\rho D_a
\psi\}$ provides an over-complete set of functions in $L^2(S^2)$.
The CSWT is given by the projection onto each wavelet basis
function in the usual manner,
\begin{equation}
(CSWT_\psi f)(a,\omega) = \int_{S^2} d\mu(\omega')(R_\omega
\psi_a)^{\ast}(\omega')f(\omega'),
\end{equation}
where $\ast$ denotes complex conjugation and
$d\mu(\omega)=\sin\theta d\theta d\varphi$ is the
rotation-invariant measure on the sphere.

\subsection{Practical Implementation}

The YAWtb team has developed the fast algorithm of CSWT and FSWT
that they integrated into the Matlab YAWtb toolbox. The scales are
discretized as
\begin{equation}
\label{disscale}
   a\in A = \{a_j \in \Re^{+} : a_j < a_{j+1} , j \in Z \},
\end{equation}
where $j$ is the scale parameter. We work on data discretized on
the equi-angular grid ${\cal G}_B$ defined by:
\begin{equation}
{\cal G}_B \equiv\{(\theta_p, \varphi_q): p,q\in {\cal Z}[2B] \},
\end{equation}
where the positive integer $B$ is the resolution parameter, ${\cal
Z}[2B] = \{0, ..., 2B-1\}$, $\theta_p = (2p+1)\pi/4B$ and
$\varphi_q = q\pi/B$. The positions are indexed by the scale
level, related to the scale in such a way that $\omega \in {\cal
G}_{B_j}$, with
\begin{equation}
\label{sphgrid}
   {\cal G}_{B_j} = \{(\theta_{jp}, \varphi_{jq})\in S^2 :\theta_{jp}=\frac{(2p+1)\pi}{4B_j}, \phi_{jq}=\frac{q\pi}{B_j}
   \},
\end{equation}
where $B_j \in {\cal Z}[2B]$.

\subsection{Spherical Wavelet Basis}

The wavelet is said to be {\it isotropic} if it does not depend on
$\varphi$ when centered at the north pole. In this work we mainly
use two different spherical mother wavelets, i.e., the isotropic
spherical Mexican hat wavelet (SMHW) and anisotropic spherical
Morlet wavelet (SMW), where anisotropic means the wavelet is not
only a function of $\theta$ but also a function of $\varphi$. We
have also performed the analysis using the difference of Gaussians
(DOG) wavelet (Bogdanova I. {\it et al.} 2004), the result of
which is almost identical to that using the SMHW. The SMHW has
been considered the best for the detection of non-Gaussian
signatures with spherical symmetry. It has already been applied to
non-Gaussian studies of the {\it COBE}-DMR data (Cay{\'o}n {\it et
al.} 2001, 2003), \emph{Planck} simulations
(Mart{\'\i}nez-Gonz{\'a}lez {\it et al.} 2002) and the {\it WMAP}
first-year data (Mukherjee \& Wang, 2004; Vielva {\it et al.}
2004; McEwen {\it et al.} 2005).
However, it is not so sensitive for the detection of anisotropic
non-Gaussianity (McEwen {\it et al.} 2005). We adopt the SMW for
the non-Gaussianity detection and the SMHW for localization. The
mother SMW wavelet is given by
%
\begin{equation}
\label{eqSMW}
   \Psi_{M}(\theta,\varphi) =
   \frac{e^{ik_0\tan(\theta/2)\cos(\varphi_0-\varphi)}
   e^{-(1/2)\tan^2(\theta/2)}}{(1+\cos\theta)^2},
\end{equation}
where $k_0$ is the projection of the wave vector of the wavelet
and $\varphi_0$ is the rotation parameter reflecting directional
features.
As defined in Mart{\'\i}nez-Gonz{\'a}lez {\it et al.} (2002), the
SMHW is given by
\begin{equation}
\label{eqSMHW}
   \Psi_{MH}(\theta,a) =
   \frac{1}{\sqrt{2\pi}N(a)}{\bigg[1+{\bigg(\frac{y}{2}\bigg)}^2\bigg]}^2
  \bigg[2 - {\bigg(\frac{y}{a}\bigg)}^2\bigg]e^{-{y}^2/2a^2},
\end{equation}
where $y\equiv 2\tan(\theta/2)$ and $N(a)\equiv a(1 + a^2/2 +
a^4/4)^{1/2}$ is a normalization constant.




\section{Non-Gaussianity Analysis}
Spherical anisotropic wavelet analysis is applied to probe the
{\it WMAP} first-year data for possible deviations from
Gaussianity. We follow a strategy similar to that of Vielva {\it
et al.} (2004) and McEwen {\it et al.} (2005), whereas our
pixelisation of the data onto the sphere and the treatment of
masks are different from theirs.

\subsection{Data Pipeline}
We adopt the same data set analyzed by Komatsu {\it et al.} 2003
and Vielva {\it et al.} 2004 in their non-Gaussianity studies. The
CMB-dominated bands (two Q-band maps at 40.7 GHz, two V-band maps
at 60.8 GHz, and four W-band maps at 93.5 GHz) are combined to
give a map with the signal-to-noise ratio enhanced. All these maps
with receiver noise, as well as beam properties, are available
from the Legacy Archive for Microwave Background Data Analysis
(LAMBDA) Web site.\footnote{http://cmbdata.gsfc.nasa.gov/} These
maps are provided in the HEALPix (G\'{o}rski {\it et al.} 1999)
format at a resolution of $N_{side} = 512$. The number of pixels
in a HEALPix map is given by $12N_{side}^2$. We project these data
in the HEALPix pixelisation onto the equi-angular spherical grid
${\cal G}_B$, where $B=256$ with the individual noise weight as
defined below, in order that the data set be suitable for matrix
computation. The surface of the sphere is redivided by the
equiangular spherical grid, and then the data are projected into
the grid according to their position and recombined within each
grid to give a weighted average in the new pixelisation. Note that
pixels in the new representation do not have equal areas, but this
is suitable for the inverse FSWT, where matrix manipulation is
used for the fast Fourier algorithm, greatly saving the
computation time. Any concern that the transformation can somehow
introduce some non-Gaussian features can be alleviated, since the
simulation has been done in the same pipeline. The transformed
data are of the size $512\times512$ pixels at a loss of resolution
that does not matter in this analysis, since very small scales are
dominated by the noise.

At a given position $\omega$ of the sky, the thermodynamic
temperature is given by
\begin{equation}
\label{eq:combination} T(\omega) = \sum_{j = 3}^{10}
{T_j}(\omega)~{w_j}(\omega),
\end{equation}
where the indices $j = 3$ and $4$ refer to the Q-band receivers,
$j = 5$ and $6$ to the ones of the V-band and $j = 7$, $8$, $9$
and $10$ to the receivers of the W-band (The indices $j = 1$ and
$2$ are used for the K and Ka receivers, respectively, which are
excluded from the analysis.) The noise weight ${w_j}(\omega)$ is
defined by
\begin{eqnarray}
\label{eq:noise} w_j(\omega) = \frac{\bar{w}_j(\omega)}{\sum_{j =
3}^{10}{\bar{w}_j}(\omega)},~~ & \bar{w}_j(\omega) =
\frac{{N_j}(\omega)}{{{\sigma_0}_j}^{2}}
\end{eqnarray}
where ${{\sigma_0}_j}$ is the standard deviation of the receiver
noise and ${N_j}(\omega)$ is the number of observations made by
the receiver $j$ at position $\omega$ on the sky (Bennett {\it et
al.} 2003a). According to equation (\ref{eq:combination}) and
(\ref{eq:noise}), the beams for each of the WMAP channels are
different and the noise variances differ from each other, which
may potentially create spatially varying beams on the sky.
However, these effects are subdominant to the scales of interest
in this paper and can be calibrated out by separate simulations of
each band.

Although the maps at selected frequencies are dominated by CMB,
Galactic foregrounds (i.e., thermal dust, free-free and
synchrotron radiations), as well as extragalactic point sources
all contribute significantly to the map. 
The WMAP team performed a foreground template fit to avoid the
Galactic emissions (Bennett {\it et al.} 2003b): the 94 GHz dust
map of Finkbeiner {\it et al.} (1999) is used as the thermal dust
template, the $H_{\alpha}$ map of Finkbeiner (2003) corrected for
extinction through the $E_{B-V}$ map of Schlegel {\it et al.}
(1998) is used as the free-free template, and finally, the
synchrotron template is the 408 MHz Haslam {\it et al.} (1982)
map. Hence, equation~(\ref{eq:combination}) is modified as
\begin{equation}
\label{eq:combination2} \hat{T}(\omega) = \sum_{j = 3}^{10}
\hat{T_j}(\omega)~{w_j}(\omega),
\end{equation}
where $\hat{T_j}(\omega)$ is the temperature at position $\omega$
for the receiver $j$ after the foreground correction.
Foreground-cleaned maps used in this analysis are available from
the LAMBDA Web site.

An independent foreground analysis of the {\it WMAP} first-year
data was performed by Tegmark {\it et al.} (2003). The Tegmark
cleaned map is constructed from a linear combination of different
band maps with the weights varying over both position and scale.
We also perform the analysis on this map in both wavelet and real
spaces.

\subsection{Mask Treatment}

Strong emissions at the Galactic plane and known radio point
sources could contaminate any intrinsic features so they should be
excluded from the map. Here we adopt the most conservative mask
called ``Kp0", which retains 76.8\% of the sky. This exclusion may
introduce edge effect to the map caused by the zero value of the
mask. Generally there are two ways to solve this edge problem. One
is to exclude the data near the edges, which has been adopted
recently (Mukherjee \& Wang 2004; Vielva {\it et al.} 2004; McEwen
{\it et al.} 2004), although any meaningful CMB features near the
edge could have been smeared out. Here we take another approach by
filling Gaussian noise in the excluded areas instead of the zero
value in order that they do not contribute to deviation from
Gaussianity. Pixel noise can be evaluated from $N_{obs}$ via
$\sigma=\sigma_{0j}/\sqrt{N_{obs}}$, where $\sigma_{0j}$ is the
standard deviation of the receiver noise of assembly $j$. Due to
the good localization nature of wavelet in both scale and position
spaces, we assume that they will not introduce significant
correlation to signals in other areas either, which have been
tested from the filtering maps such that the filtered cold and hot
spots are not caused by particular masks and there is no
correlation between these spots and the point sources. Note that
the non-Gaussian detection is quite insensitive to the particular
choice of the exclusion masks (Mukherjee \& Wang, 2004; Vielva
{\it et al.} 2004; McEwen {\it et al.} 2005). This approach should
not introduce non-Gaussian signals, especially at the scales we
are interested in this work. The final preprocessed Q-V-W-combined
WMAP team map and the Tegmark cleaned map are shown in Fig.~1.

\begin{figure}
    \centering\subfigure[preprocessed combined WMAP team map]
        {\includegraphics[scale=0.39]{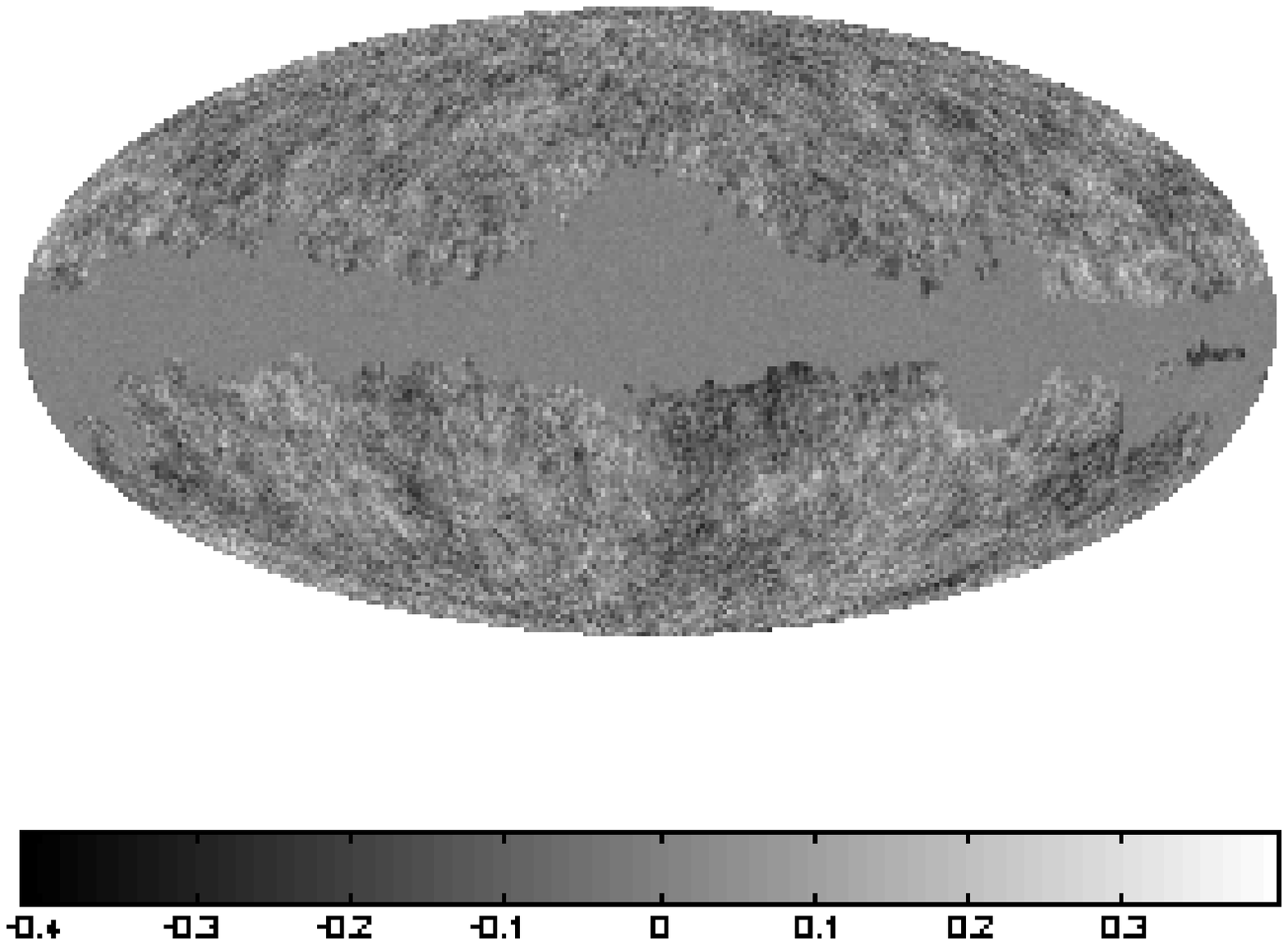}}
    \centering\subfigure[preprocessed Tegmark cleaned map]
        {\includegraphics[scale=0.39]{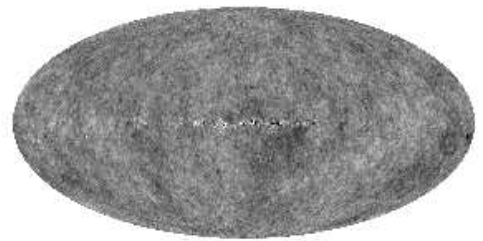}}
    \caption{CMB temperature anisotropy maps preprocessed
    from the WMAP team map and the Tegmark cleaned map to
    be analyzed in this non-Gaussianity study. 
All maps
     in this paper are plotted in Galactic coordinates with
     the Galactic center $(l,b)=(0,0)$ in the middle and Galactic
     longitude $l$ increasing to the left.}
\end{figure}

\subsection{Non-Gaussianity Statistics}

The wavelet transform is a linear operation, hence the wavelet
coefficients of a Gaussian signal will also obey a Gaussian
distribution. So we can probe a full-sky map for non-Gaussianity
either by looking for deviations from Gaussianity in wavelet space
or in real space. We thus examine non-Gaussian features in wavelet
space and then localize them in real space. In wavelet space the
localization can only be done at a specific scale, whereas in real
space using the inverse framed wavelet transform technique, one
can localize any deviations of Gaussianity within some scale
ranges that would be more natural for the real signals. This is
because a map obtained from the inverse framed wavelet transform
for a range of scales is equivalent to filtering the original map
with a band-pass filter; normally artifacts or spurious
oscillations will be produced in a filtered map if the bandwidth
is too narrow.

At a given scale ($a$), we use the third and fourth moments about
the mean as non-Gaussian estimators for the test, i.e., skewness
($S(a)$) and excess kurtosis ($K(a)$) given by:
\begin{eqnarray}
S(a) & = & \frac{1}{N_a} \sum_{i = 1}^{N_a} {w_{i}(a)}^3  {\bigg
/} {\sigma(a)}^3
\\
K(a) & = & \frac{1}{N_a} \sum_{i = 1}^{N_a} {w_{i}(a)}^4  {\bigg
/} {\sigma(a)}^4 - 3,
\end{eqnarray}
where $N_a$ is the number of coefficients and $\sigma(a)$ is the standard deviation of
the wavelet coefficients on
scale $a$.

For a Gaussian distribution, the statistics $S(a)$ and $K(a)$ have
zero mean value at each scale $a$. Thus deviations from zero value
in these statistics will indicate the existence of non-Gaussianity
in the spherical wavelet coefficients and hence in the
corresponding real map.

In the wavelet space we perform the CWT at scales $a \in [0.02,
0.05]$ corresponding to about ($1^{\circ}, 30^{\circ}$) on the
sky, as well as within a range of directions $\varphi_0 \in
[0^{\circ}, 360^{\circ})$ with a $30^{\circ}$ interval. 
To localize any expected non-Gaussian features, we perform the
FSWT (\texttt{fwtsph.m} in the YAWtb) and the inverse FSWT
(\texttt{ifwtsph.m} in the YAWtb) in order to filter the signal
through some ranges of scales rather than only at a certain scale.
Scales larger than about $3^{\circ}$ on the sky are selected
because the previous tests in wavelet space only found
non-Gaussianity on such scales (Mukherjee \& Wang 2004; Vielva
{\it et al.} 2004; McEwen {\it et al.} 2005). This is the first
spherical inverse wavelet analysis applied to the CMB data in
which the frame technique is essential.




\subsection{Monte Carlo Simulations}

Monte Carlo simulations are performed to construct confidence
levels for the non-Gaussianity test statistics described above.
First, using CMBFAST (Seljak \& Zaldarriaga 1996), we calculated
the power spectrum $C_l$ using the cosmological parameters
estimated by the WMAP team (Spergel {\it et al.} 2003). Second,
10,000 Gaussian CMB realizations are produced and convolved at
each of the WMAP receivers with beam functions. Third, we
transform the simulated data from harmonic to real space and add
Gaussian noise according to the number of observations per pixel
and the noise standard deviation per observation. Finally, all the
maps from the eight receivers are combined following equation
(\ref{eq:combination2}). Monte Carlo simulations of beams and the
foreground where the CMB signal is almost negligible are also
performed separately in a similar way.

Since the weights used to construct the Tegmark cleaned map differ
from those used by the {\it WMAP} team, we should carry out other
Gaussian simulations strictly following the Tegmark construction
method. However, there is no simple way of estimating the noise in
each pixel, since much of the noise is due to residual
foregrounds, which vary
strongly across the sky and are thus correlated between pixels. 
Thus, Gaussian simulations of the Q-V-W-combined map are also used
for the Tegmark map, since for both of the two maps, the weights
sum to unity, and we assume that the slight difference in the
linear combination of maps should not cause significant changes in
the Gaussian confidence levels (McEwen {\it et al.} 2005).

\section{Results}

Analysis in wavelet space has been performed on both of the
Q-V-W-combined {\it WMAP} map (here and throughout, all CMB maps
used are foreground-removed maps) and the Tegmark cleaned map to
test the non-Gaussianity in the {\it WMAP} first-year data. We
also investigate the possible sources of the detected
non-Gaussianity. Real space analysis is performed to localize
possible deviations over some ranges of scales on the sky.
\begin{figure}
    \centering\subfigure[skewness S(a)]
        {\includegraphics[scale=0.39]{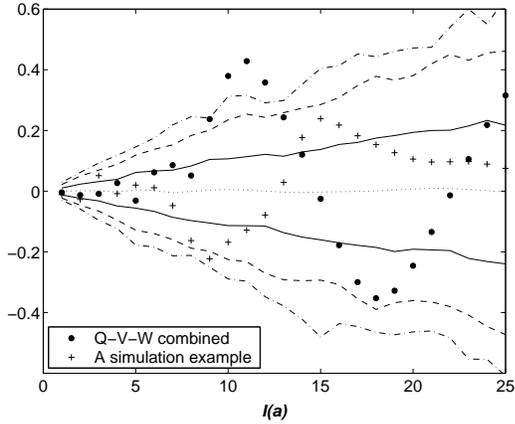}}
    \centering\subfigure[kurtosis K(a)]
        {\includegraphics[scale=0.39]{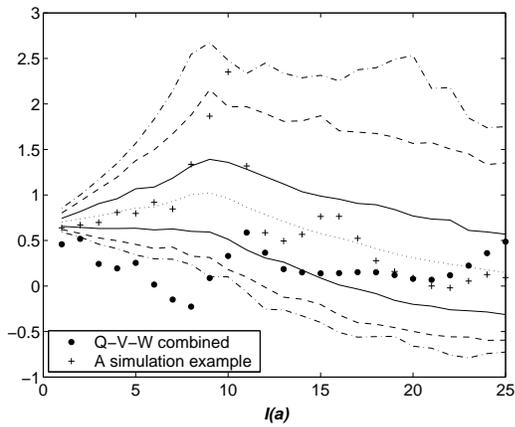}}
    \caption{Skewness and kurtosis values of the SMW analysis convolved
     with the Q-V-W combined WMAP map are shown as filled circles. Here and
     throughout, the acceptance intervals at the 68\% ({\it solid lines}), 95\% ({\it dashed lines})
     and 99\% ({\it dash-dotted lines}) significance levels given by 10,000 Monte Carlo
    simulations are shown. Note that the error bars are too
    small compared to the coefficients values and can be ignored in
    the figures. The values from one example of the simulated
    Gaussian maps ({\it plus sign}s) are also shown for comparison.
    Index $I(a)\equiv 50a$ and the corresponding size on the sky can be obtained
    from Eqn. (\ref{size}), where scale index $1$ corresponds to $1.15^{\circ}$. Note that
    only the values obtained from the azimuthal orientation $\varphi_0 \sim 0^{\circ}$ (i.e., corresponding to the
    maximum deviations from Gaussianity) are shown.}
\end{figure}
\begin{deluxetable}{rrrrrrr} \tablecolumns{7}
\tablecaption{Deviations and significance levels of spherical
wavelet coefficient statistics obtained from analyzing the Q-V-W
combined WMAP map. Significant levels are calculated from 10,000
Gaussian simulations. The number of standard deviations the
observation deviates from the mean is given by $N_{\sigma}$; the
corresponding significance level of the detected non-Gaussianity
is given by $\delta$.\label{tableskandku}} \tablehead{
\multicolumn{3}{c}{Skewness} & \colhead{}
& \multicolumn{3}{c}{Kurtosis} \\
\cline{1-3} \cline{5-7} \\
\colhead{I(a)} & \colhead{$N_{\sigma}$}   & \colhead{$\delta$} &
\colhead{} & \colhead{I(a)} & \colhead{$N_{\sigma}$} &
\colhead{$\delta$}} \startdata
10 & 3.31 & 99.88\% & & 3 & -3.77 & 99.88\% \\
11 & 3.49 & 99.79\% & & 4 & -3.45 & 99.65\% \\
12 & 2.84 & 99.61\% & & 6 & -3.17 & 98.90\% \\
   &      &         & & 7 & -3.38 & 99.21\% \\
   &      &         & & 8 & -3.01 & 98.52\% \\
\enddata
\end{deluxetable}

\subsection{Spherical Wavelet Coefficient Statistics}

For the Q-V-W-combined {\it WMAP} map, the skewness and kurtosis
of the wavelet coefficients at different scales are illustrated in
Fig.~2; only the values obtained from the orientation
corresponding to the maximum deviations from Gaussianity are
shown. Deviations from Gaussianity are detected in both the
skewness and kurtosis of the combined map. The numbers of
deviations and the corresponding significance levels obtained from
10,000 Gaussian simulations are displayed in Table
(\ref{tableskandku}). Since we have tested that deviations using
SMW in certain directions are larger than those using SMHW (McEwen
{\it et al.} 2005), we only present the SMW results here. In the
skewness, deviations are detected on scales $a_{10}\sim
11^{\circ}.42$ to $a_{12}\sim 13^{\circ}.69$, with a maximum
around the scale $a_{11}\sim 12^{\circ}.55$ at a significance
level of $>99\%$. In the kurtosis, deviations are also detected on
scales $<a_{9}\sim 10^{\circ}.29$ with a maximum around the scale
$a_{8}\sim 9^{\circ}.14$, showing some consistency with the
kurtosis test in Vielva {\it et al.} (2004). Note that $a_8$,
$a_{10}$, $a_{11}$, and $a_{12}$ are scale parameters defined in
equation (\ref{disscale}) that are related to physical scales
according to equation (\ref{size}).

\subsection{Non-Gaussianity Localization}

Here we only present results obtained from the isotropic SMHW
analysis in real space, since anisotropic SMW will focus only on
some signals at a certain direction on the sky. After filtering
the map through certain scales, the $3$ $\sigma$ threshold filter
bank is performed so that only the values of which the absolutes
are larger than the threshold value are preserved. Results of this
analysis for both the Q-V-W-combined {\it WMAP} and the Tegmark
cleaned maps are shown in Figs.~3 and 4. Coordinates of the
filtered non-Gaussian spots in both maps and their corresponding
numbers of deviations are listed in Table (\ref{tablecoord}). Our
preliminary work does not reveal any obvious correlation with
known sources, although currently we cannot exclude this
possibility. Clearly, extensive further work is required.
Meanwhile, the filtering results provide some clues for unknown
foreground components, which is another aim of our work.
Some of the spots are only detected either in the combined {\it
WMAP} map or in the Tegmark cleaned map, and some spots are seen
in both maps, i.e., spots 3, 4, and 7. Note that all the spots
found in the Tegmark map but absent in the {\it WMAP} combined
map, i.e., spots 12$\sim$17, are near the Galactic plane. They
should have been masked out by the Kp0 mask in the WMAP combined
map even if they do exist. On the other hand, all the spots found
in the Tegmark map at high latitudes, i.e., spots 3, 4 and 7, are
all found in the {\it WMAP} combined map, whereas there are other
high-latitude spots found in the combined map but absent in the
Tegmark map, which we discuss later in the source determination
that the Tegmark map seems to be ``cleaner'' than the {\it WMAP}
combined map at some level. Indeed, we should be very cautious of
these spots, especially those near the Galactic plane. Whether
they are true or not needs further investigation. We suggest that
the spots at high latitudes are more real. These
different-filtering spots may have something to do with varying
noise weights used, or with foreground residuals, since the two
maps have been processed using different foreground removal
techniques. However, we discuss the possible sources of this
non-Gaussianity in detail below. We have carried out the analysis
after the spots are removed from the maps, and there is indeed no
significant deviation from Gaussian fluctuations, indicating these
spots do contribute to the detected non-Gaussianity. Note that
spot $7$ corresponds to the cold spot pointed out in Vielva {\it
et al.} (2004) and further discussed in Cruz {\it et al.} (2005);
the deep hole corresponding to spots $15$, $16$, and $17$ is also
present in the {\it WMAP} internal linear combination map and does
not fail a Minkowski functional analysis (Colley \& Gott 2003) in
which some non-Gaussian features are more sensitive in wavelet
spaces than in others.

\begin{figure}
    \centering\subfigure[The filtered combined WMAP map]
        {\includegraphics[scale=0.39]{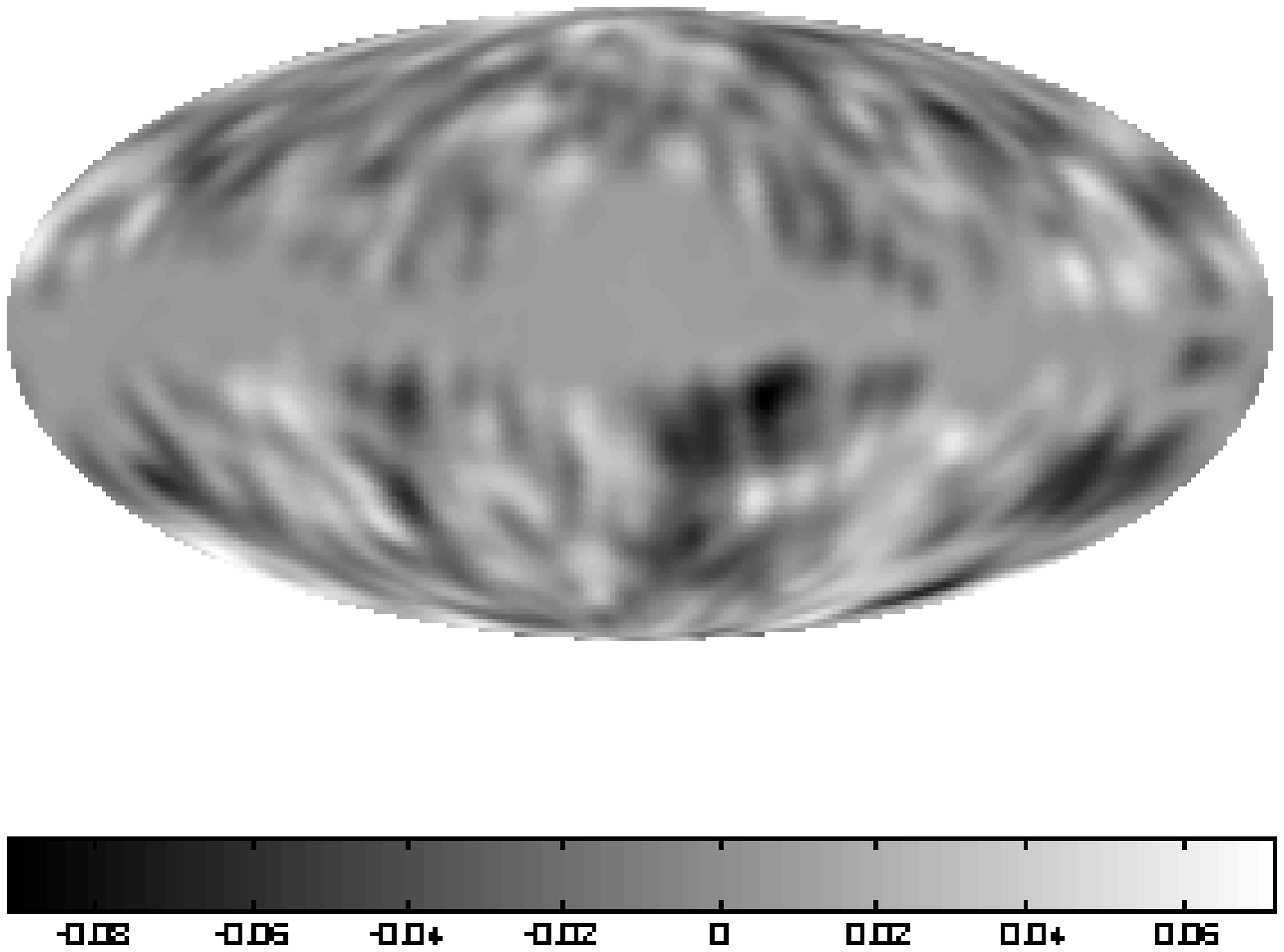}}
    \centering\subfigure[After $3\sigma$ thresholded]
        {\includegraphics[scale=0.39]{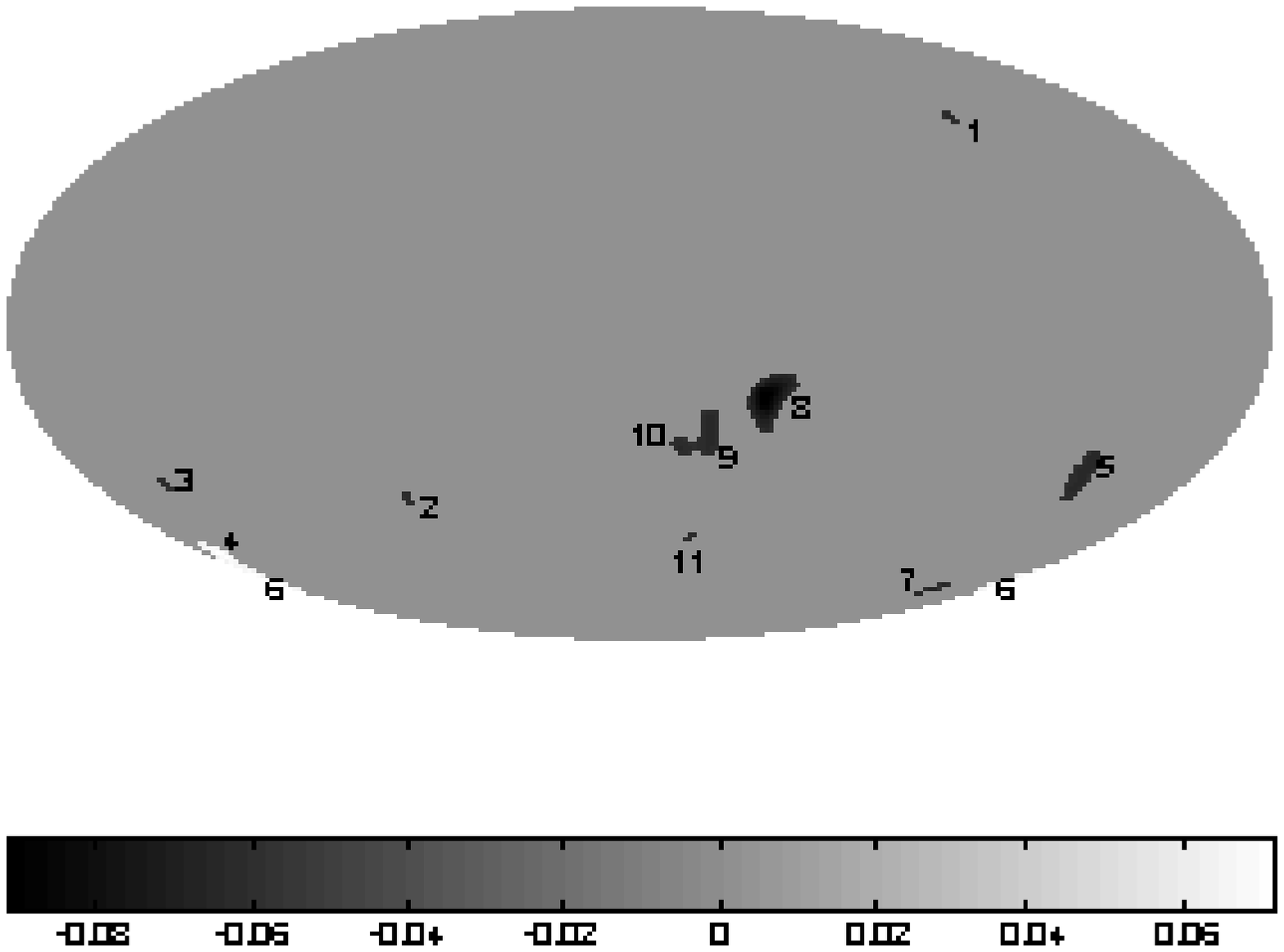}}
    \caption{Inverse spherical wavelet transformed and $3$ $\sigma$
    thresholded Q-V-W-combined WMAP maps in real space.}
\end{figure}
\begin{figure}
    \centering\subfigure[The filtered Tegmark map]
        {\includegraphics[scale=0.39]{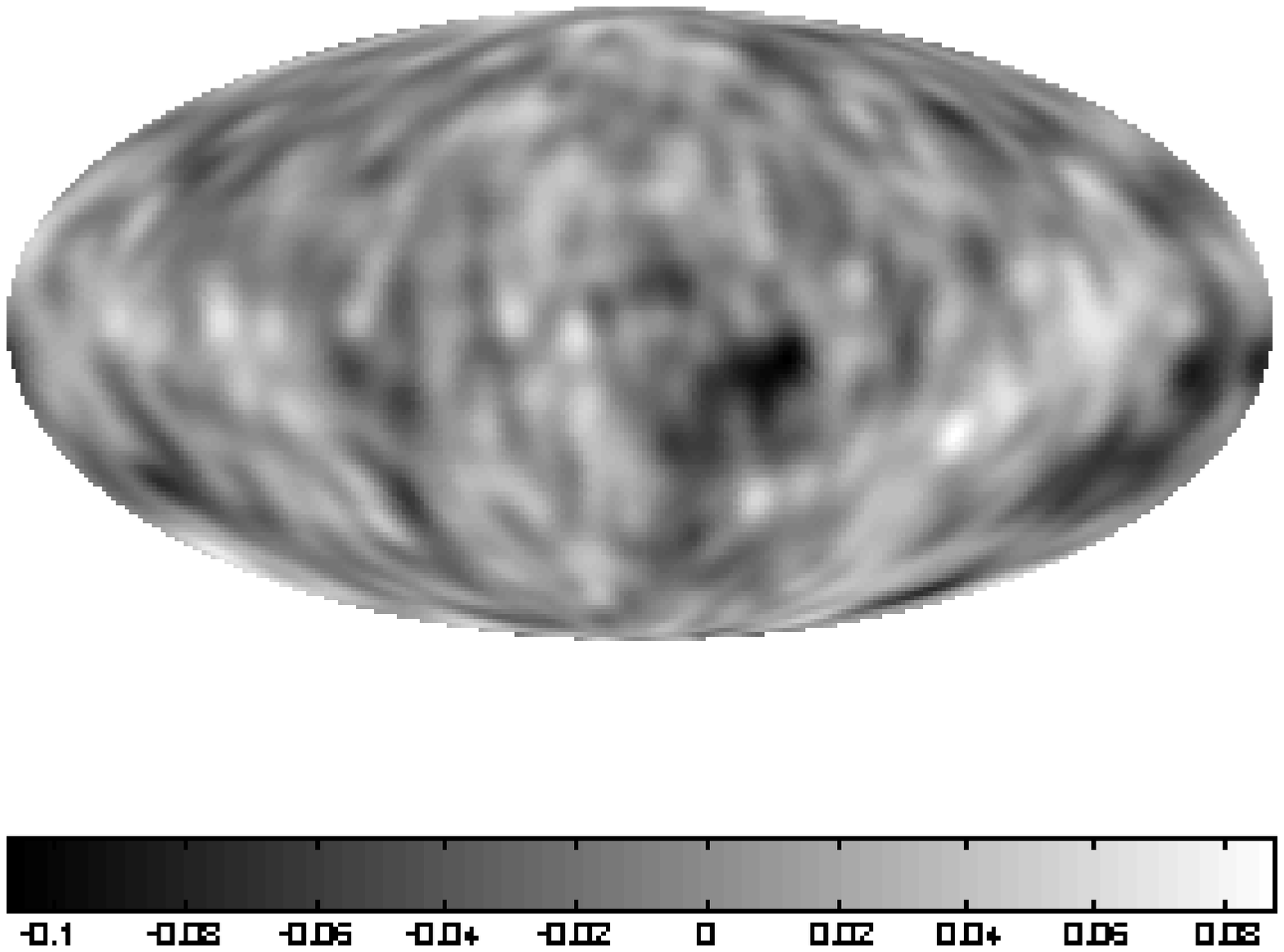}}
    \centering\subfigure[After $3\sigma$ thresholded]
        {\includegraphics[scale=0.39]{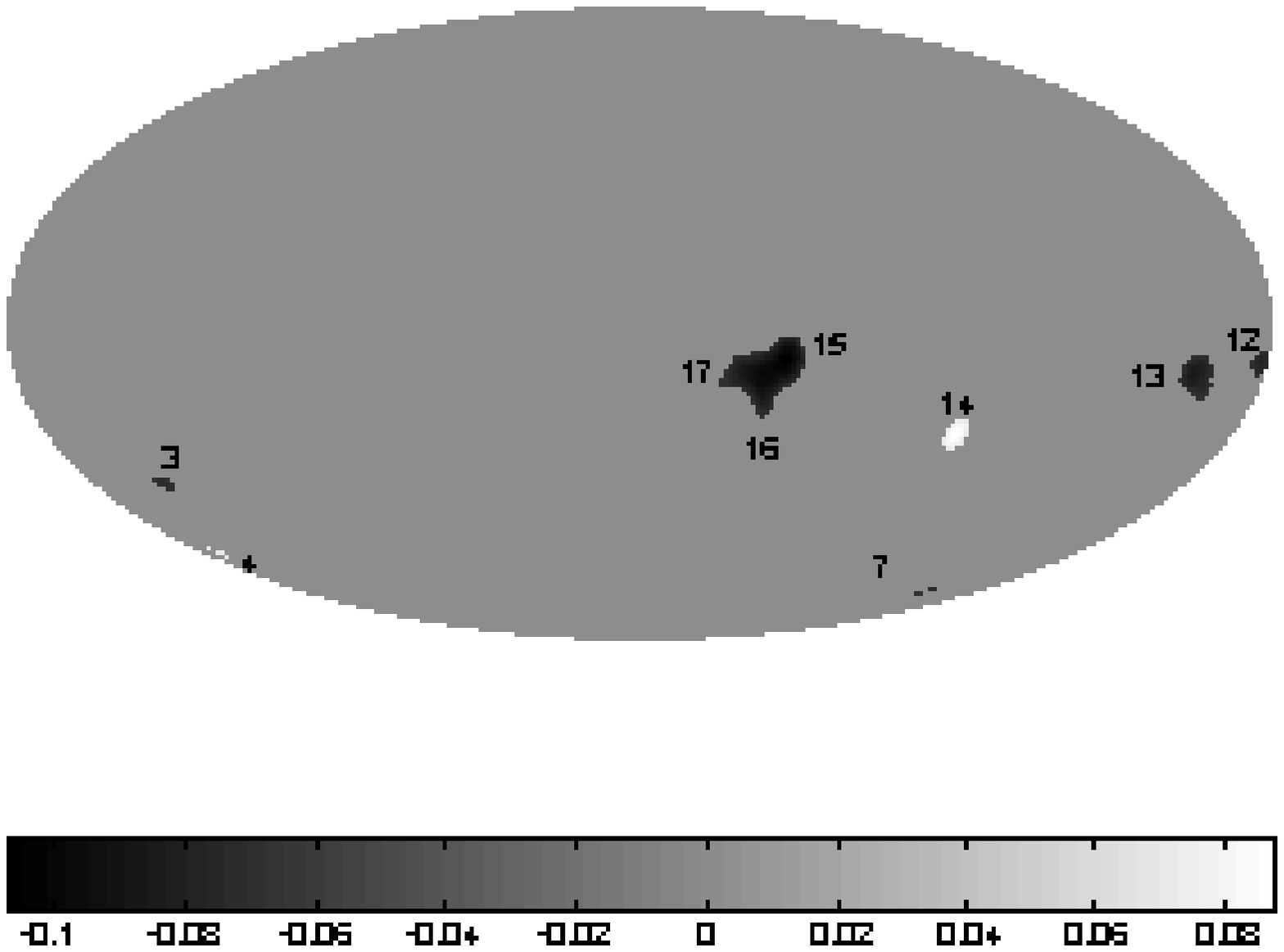}}
    \caption{Inverse spherical wavelet transformed and $3$ $\sigma$
    thresholded Tegmark cleaned maps in real space.}
\end{figure}
\begin{deluxetable}{rrrrrrrr} \tablecolumns{8}
\tablecaption{Coordinates of the filtered non-Gaussian spots. The
number of standard deviations the spot deviate from the mean are
given by $N_{\sigma}$. The space means the spot is not detected in
the specific map.\label{tablecoord}} \tablehead{ \colhead{} &
\multicolumn{3}{c}{The WMAP combined} & \colhead{}
& \multicolumn{3}{c}{The Tegmark cleaned} \\
\cline{2-4} \cline{6-8} \\
\colhead{No.}    & \colhead{$b$}   & \colhead{$l$}  &
\colhead{$N_{\sigma}$} & \colhead{}      & \colhead{$b$}  &
\colhead{$l$}          & \colhead{$N_{\sigma}$}} \startdata
1 & $40^{\circ}$   & $245^{\circ}$ & -3.12 & &               &               &       \\
2 & $-34^{\circ}$  & $79^{\circ}$  & -3.06 & &               &               &       \\
3 & $-31^{\circ}$  & $157^{\circ}$ & -3.08 & & $-31^{\circ}$ & $157^{\circ}$ & -3.11 \\
4 & $-47^{\circ}$  & $174^{\circ}$ &  3.20 & & $-47^{\circ}$ & $174^{\circ}$ & 3.06 \\
5 & $-29^{\circ}$  & $218^{\circ}$ & -3.15 & &               &               &       \\
6 & $-54^{\circ}$  & $180^{\circ}$ &  3.28 & &               &               &       \\
7 & $-58^{\circ}$  & $210^{\circ}$ & -3.23 & & $-58^{\circ}$ & $210^{\circ}$ & -3.09 \\
8 & $-14^{\circ}$  & $322^{\circ}$ & -3.69 & &               &               &       \\
9 & $-20^{\circ}$  & $340^{\circ}$ & -3.14 & &               &               &       \\
10 & $-23^{\circ}$ & $348^{\circ}$ & -3.09 & &               &               &       \\
11 & $-43^{\circ}$ & $342^{\circ}$ & -3.06 & &               &               &       \\
12 &               &               &       & & $-8^{\circ}$  & $182^{\circ}$ & -3.28  \\
13 &               &               &       & & $-10^{\circ}$ & $199^{\circ}$ & -3.35 \\
14 &               &               &       & & $-21^{\circ}$ & $264^{\circ}$ & 3.38 \\
15 &               &               &       & & $-7^{\circ}$  & $318^{\circ}$ & -4.01 \\
16 &               &               &       & & $-11^{\circ}$ & $323^{\circ}$ & -3.76 \\
17 &               &               &       & & $-10^{\circ}$ & $331^{\circ}$ & -3.49 \\
\enddata
\end{deluxetable}
\begin{figure*}
    \centering\subfigure[combined Q1Q2]
        {\includegraphics[scale=0.3]{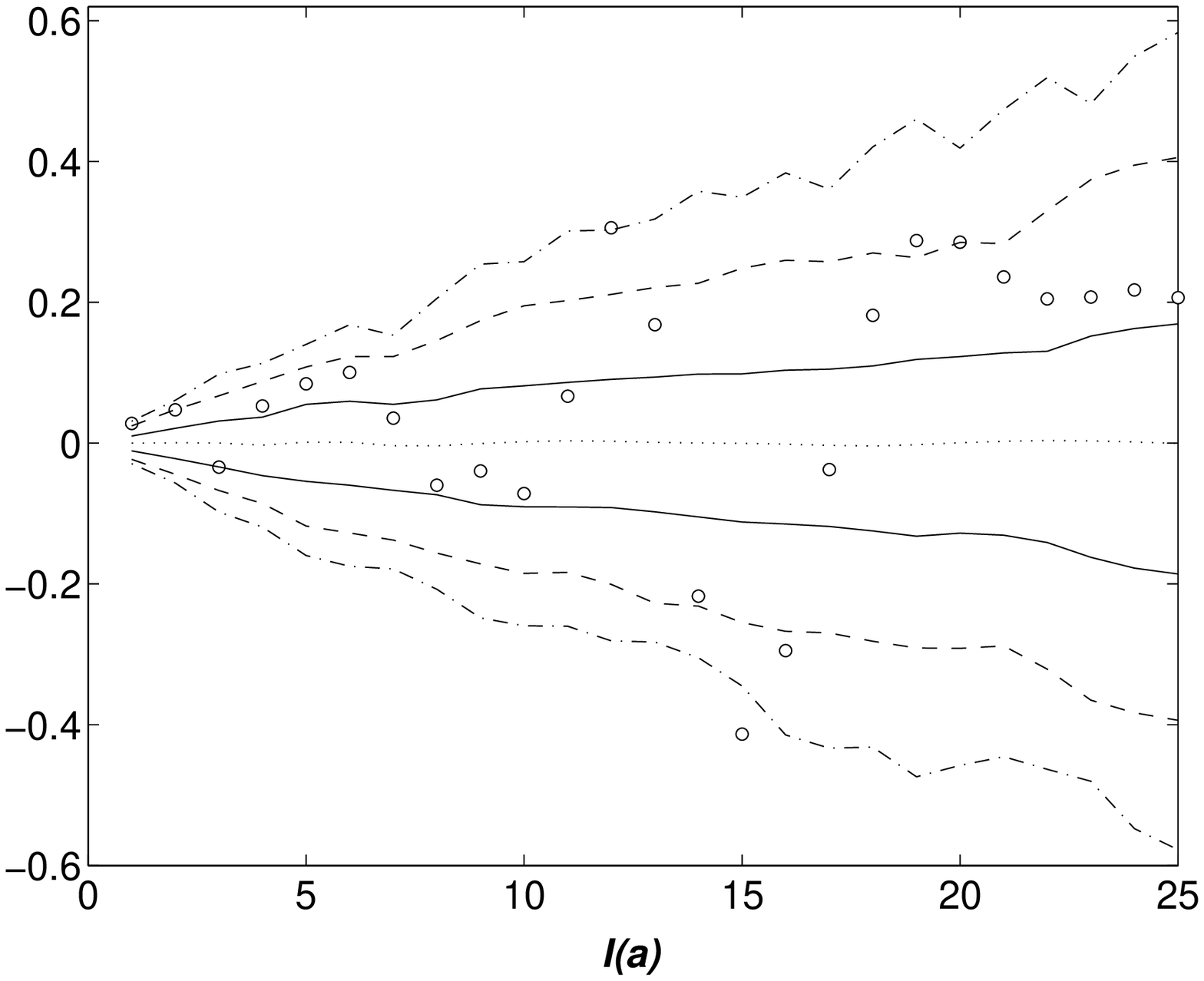}}
    \centering\subfigure[combined V1V2]
        {\includegraphics[scale=0.3]{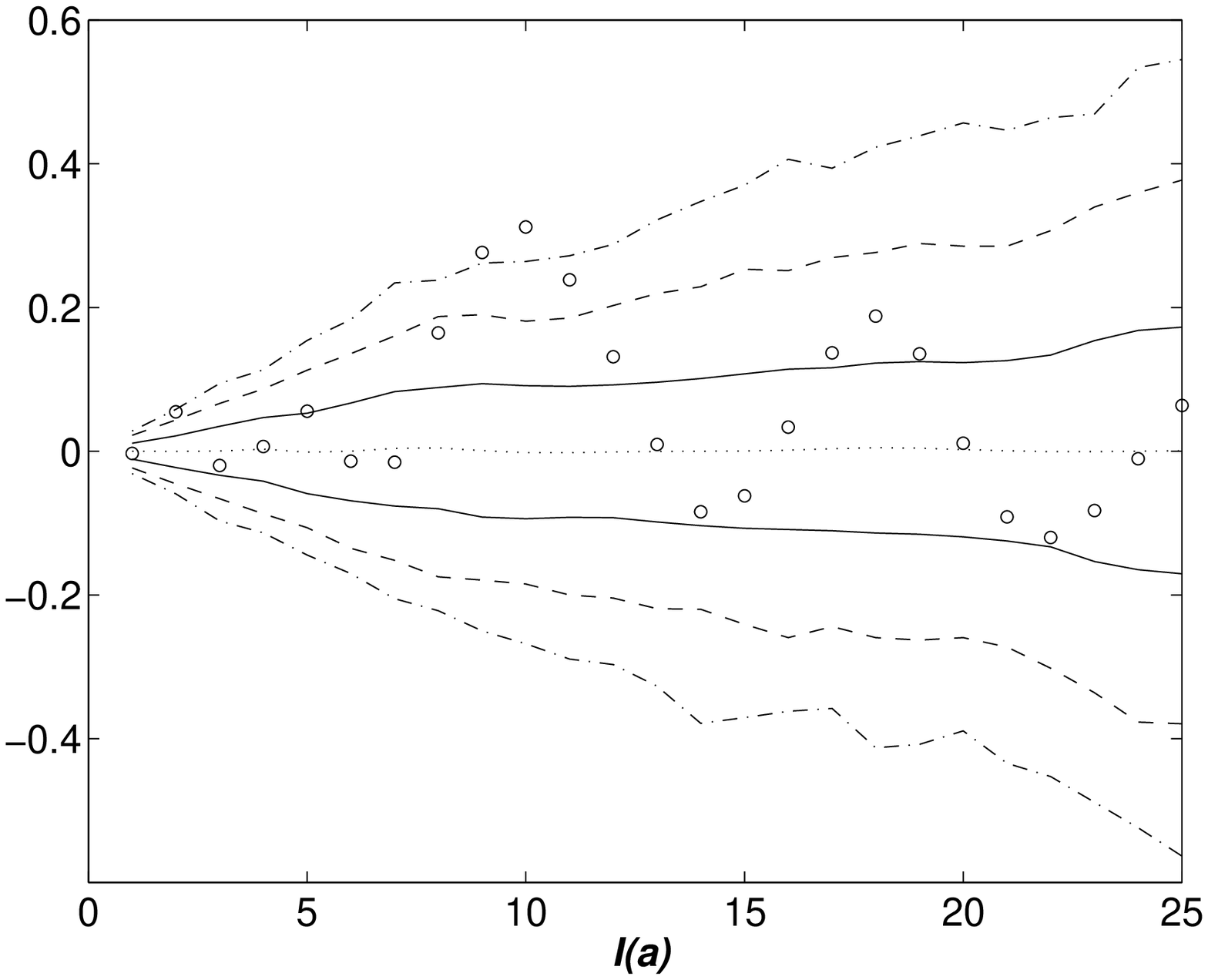}}
    \centering\subfigure[combined W1W2+W3W4]
        {\includegraphics[scale=0.3]{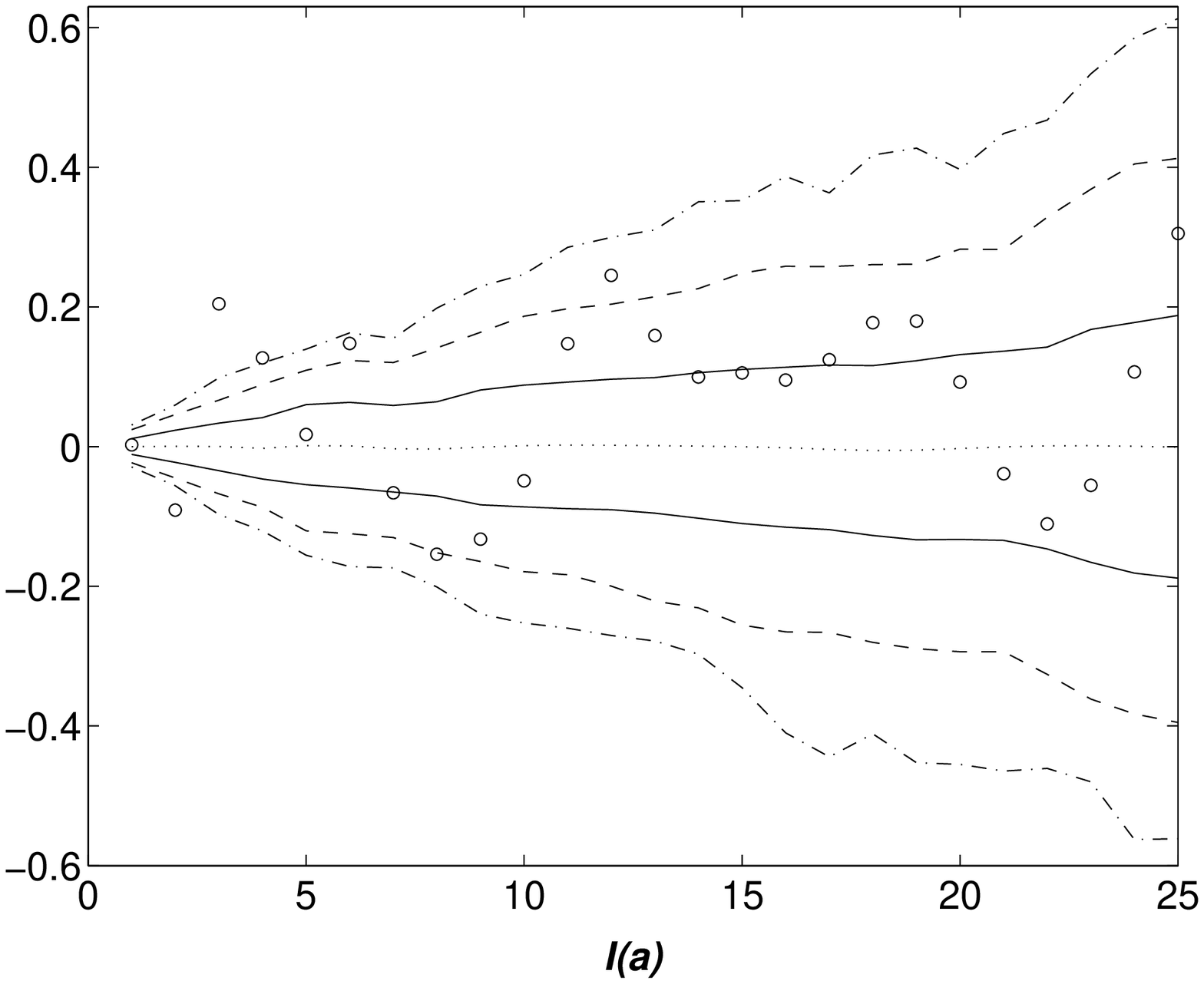}}
    \caption{Skewness values obtained from the analysis of different
    receiver combinations, where the CMB and foreground contributions
    are negligible, shown as circles. Note that they all seem to be
    compatible with Gaussian simulations at the scales at which the non-Gaussianity
    is detected and exhibit very different patterns from those of the CMB maps,
    indicating that systematic effects due to beams do not significantly contribute to the
    detected non-Gaussianity.}
\end{figure*}

\subsection{Sources of Non-Gaussianity}

We perform analysis in the wavelet space using SMW to discuss the
sources of the detected non-Gaussianity. Vielva {\it et al.}
(2004) have studied this issue using isotropic SMHW and concluded
that systematics, foregrounds, and noise can be rejected as the
source, except for possible intrinsic fluctuations. Here we
re-examine the possible sources of the detected non-Gaussianity.

We subtract maps of the receivers at the same frequency to produce
maps almost free from CMB and foreground. These maps include
possible systematic instrumental features like the beams and noise
of each band. SMW analysis has been performed, and the skewnesses
with confidence levels obtained from individual Monte Carlo
simulations of each band are shown in Fig.~5. They do not show
significant non-Gaussian fluctuations and exhibit completely
different patterns from the skewness curve obtained from the
Q-V-W-combined map.
Note that Q -band has 99\% outliers at $\it{I(a)}$$\sim 12$ and
$\sim 15$, V -band at $\sim 10$, and W -band at $\sim 3$ and $\sim
4$. Thus, both Q and V channels show some deviations around the
scales where the non-Gaussianity is detected. However, they are
not significant enough to explain the detected deviations, which
achieve their maxima around $\it{I(a)}$$\sim 11$ in all three
bands. The numbers of deviations and the corresponding
significance levels are displayed in Table (\ref{tablenb}). In
addition, we have tested this by adding an overestimated noise
contamination supplied by the WMAP team to Gaussian CMB
simulations, but the detected non-Gaussian signal does not appear
in the analysis.

We also examine each receiver by performing the analysis on the
Q1, Q2, V1, V2, W1, W2, W3, and W4 maps. The skewnesses with
confidence levels from Gaussian simulations are shown in
Fig.~6(a). They all exhibit a  pattern similar to the combined
{\it WMAP} map; thus, the deviations are not caused by any
particular receiver. These tests indicate that systematic effects
do not contribute significantly to the detected non-Gaussianity.

For the possible contribution of foreground to the detected
non-Gaussianity, we note that it can be categorized into two
parts: one is from the foreground emission templates adopted by
the {\it WMAP} team and another is some possible residual
contamination after foreground-template correction. We check the
former by (1) analyzing a map almost free of CMB signals; the map
is made by subtracting the two receivers of Q band and the two
receivers of V band from the four receivers of W band. This map
will contain significant contributions from foregrounds and noise
derived from the WMAP data (Vielva {\it et al.} 2004). The
skewness of this map and its own confidence levels from Monte
Carlo simulations are shown in Fig.~7(a), showing very different
patterns of that from the cleaned map ( Fig.~6(a)), and no
significant deviations from Gaussian fluctuations are seen. (2) We
also analyze the statistical properties of the foreground
templates obtained by the {\it WMAP} team by adding overestimated
foreground contaminations to a simulated Gaussian CMB map. The
skewness of these maps and their corresponding confidence levels
are shown in Fig.~7(b), where the detected non-Gaussian signal
does not appear, indicating no correlation with Galactic
foregrounds. In addition, since the non-Gaussian signals are
detected at intermediate scales, the foreground emission templates
can be excluded as the source of the detected non-Gaussianity.

\begin{deluxetable}{rrrrrrrrrrr} \tablecolumns{11}
\tablecaption{Deviations and significance levels of several
outlying spherical wavelet coefficient statistics obtained from
analyzing different receiver combinations, where the CMB and
foreground contributions are negligible. Significant levels are
calculated from 10,000 Gaussian simulations. The number of
standard deviations the observation deviates from the mean is
given by $N_{\sigma}$; the corresponding significance level of the
detected non-Gaussianity is given by $\delta$. \label{tablenb}}
\tablehead{\multicolumn{3}{c}{Q1Q2} & \colhead{} &
\multicolumn{3}{c}{V1V2} & \colhead{}
& \multicolumn{3}{c}{W1W2+W3W4} \\
\cline{1-3} \cline{5-7} \cline{9-11} \\
\colhead{I(a)} & \colhead{$N_{\sigma}$}   & \colhead{$\delta$} &
\colhead{} & \colhead{I(a)} & \colhead{$N_{\sigma}$} &
\colhead{$\delta$} & \colhead{} & \colhead{I(a)} &
\colhead{$N_{\sigma}$} & \colhead{$\delta$}} \startdata
12 &  3.0 & 99.2\% & &  9 & 3.0 & 99.4\% & &  2 & -4.0 &  99.9\%  \\
15 & -3.3 & 99.5\% & & 10 & 3.4 & 99.6\% & &  3 &  5.8 & $>$99.9\%  \\
   &       &         & &    &      &     & &  4 &  2.9 &  99.3\%  \\
\enddata
\end{deluxetable}
\begin{figure*}
    \centering\subfigure[each receiver and the combined]
    {\includegraphics[scale=0.39]{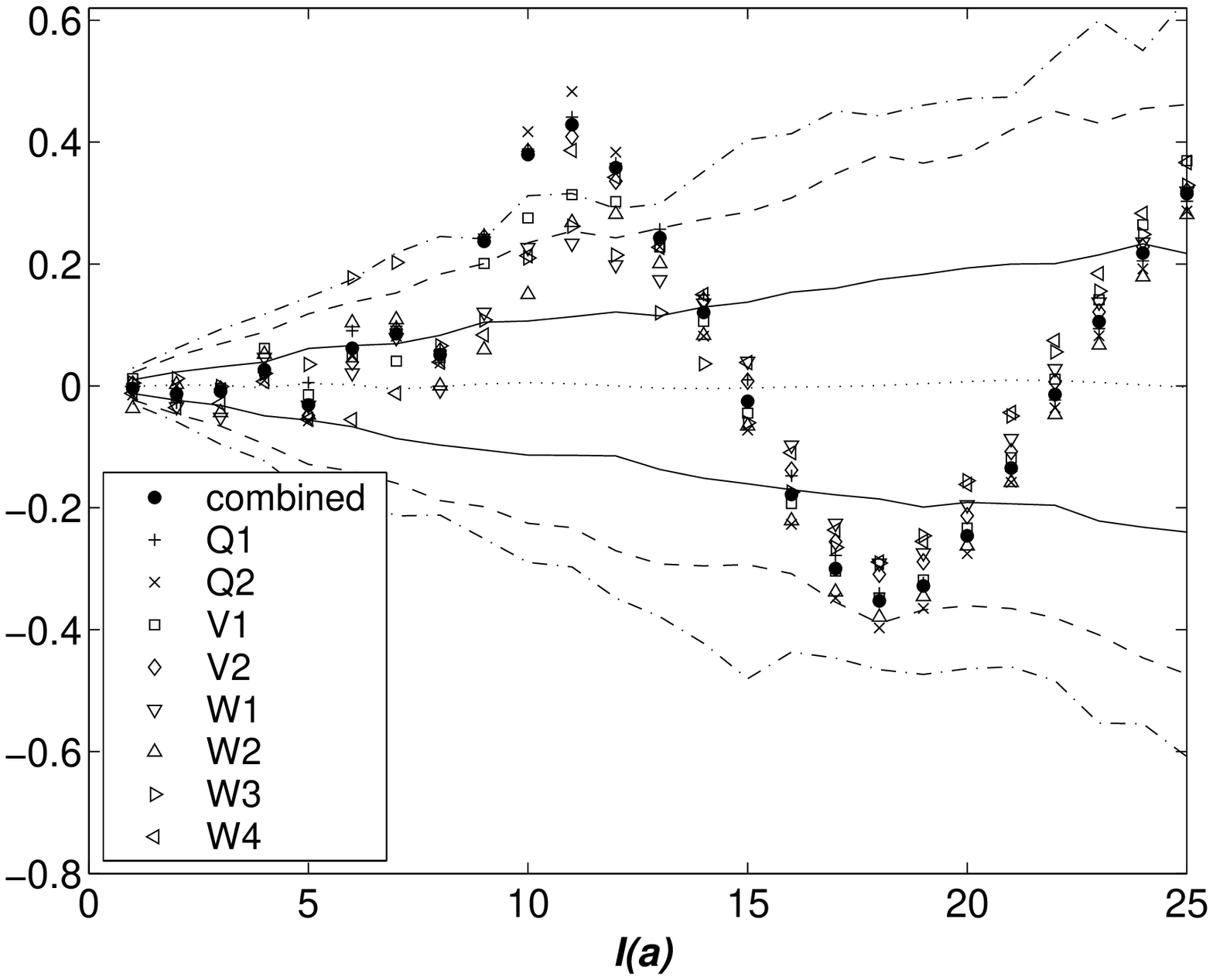}}
    \centering\subfigure[the Q-band]
    {\includegraphics[scale=0.39]{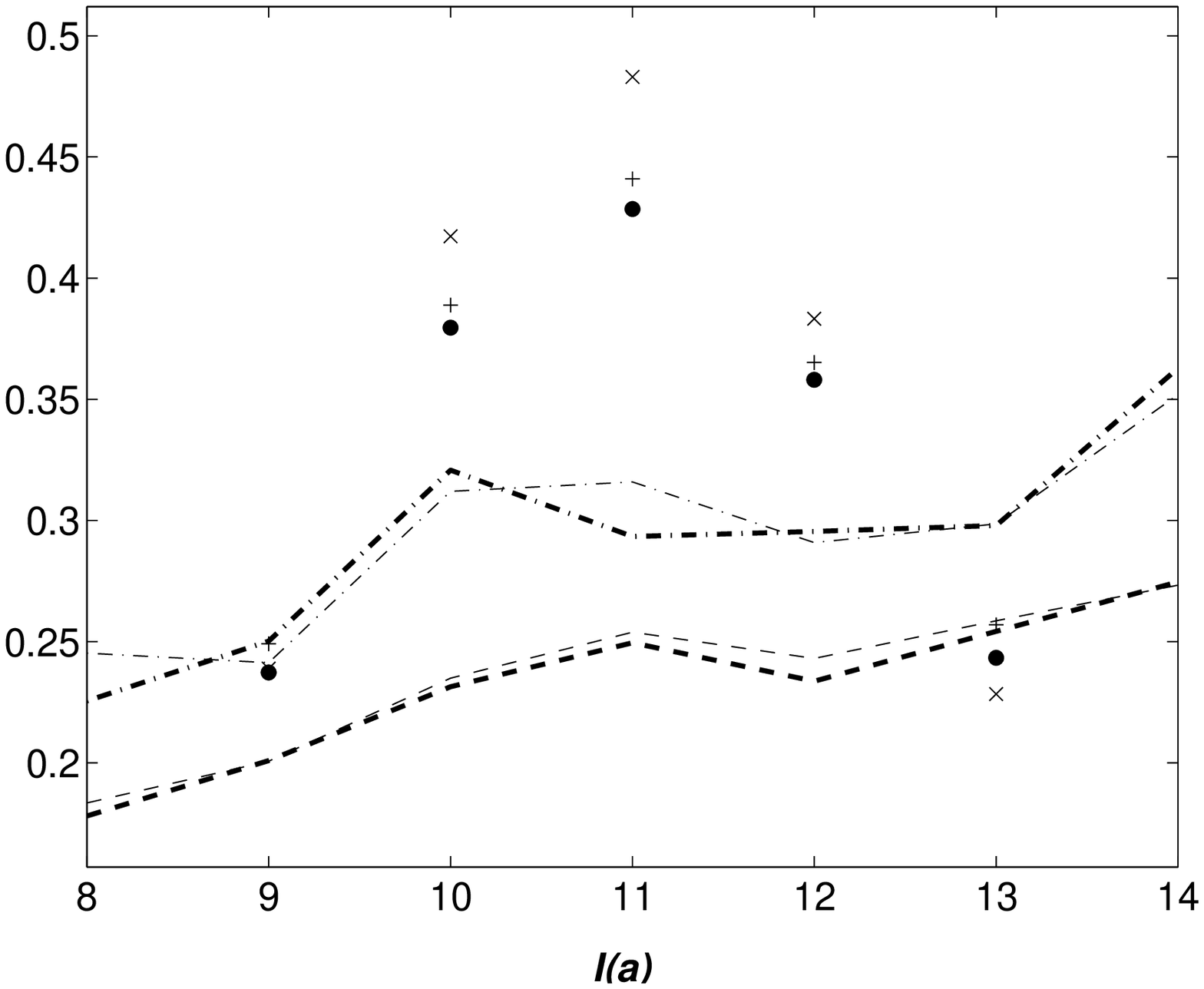}}
    \centering\subfigure[the V-band]
    {\includegraphics[scale=0.39]{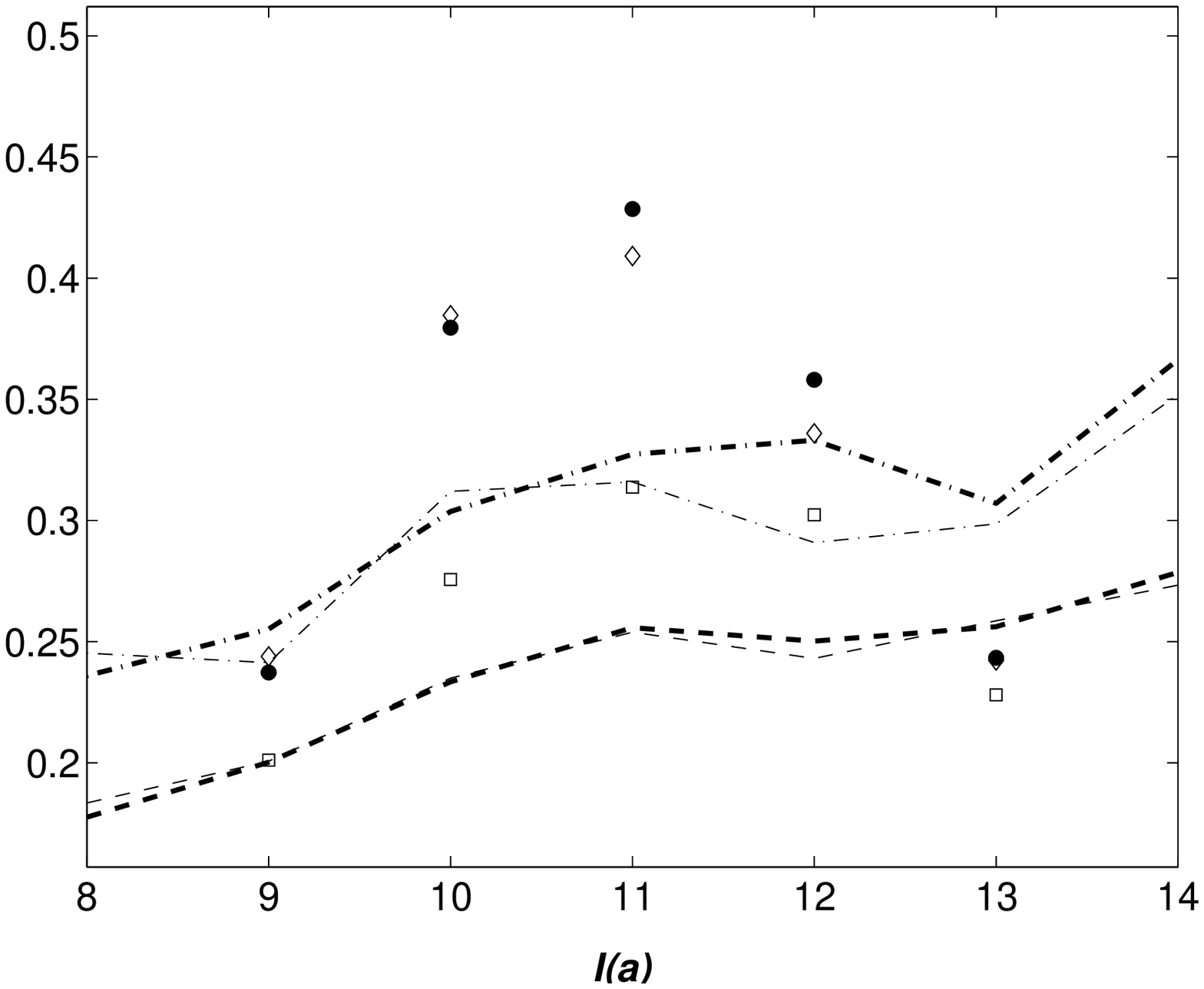}}
    \centering\subfigure[the W-band]
    {\includegraphics[scale=0.39]{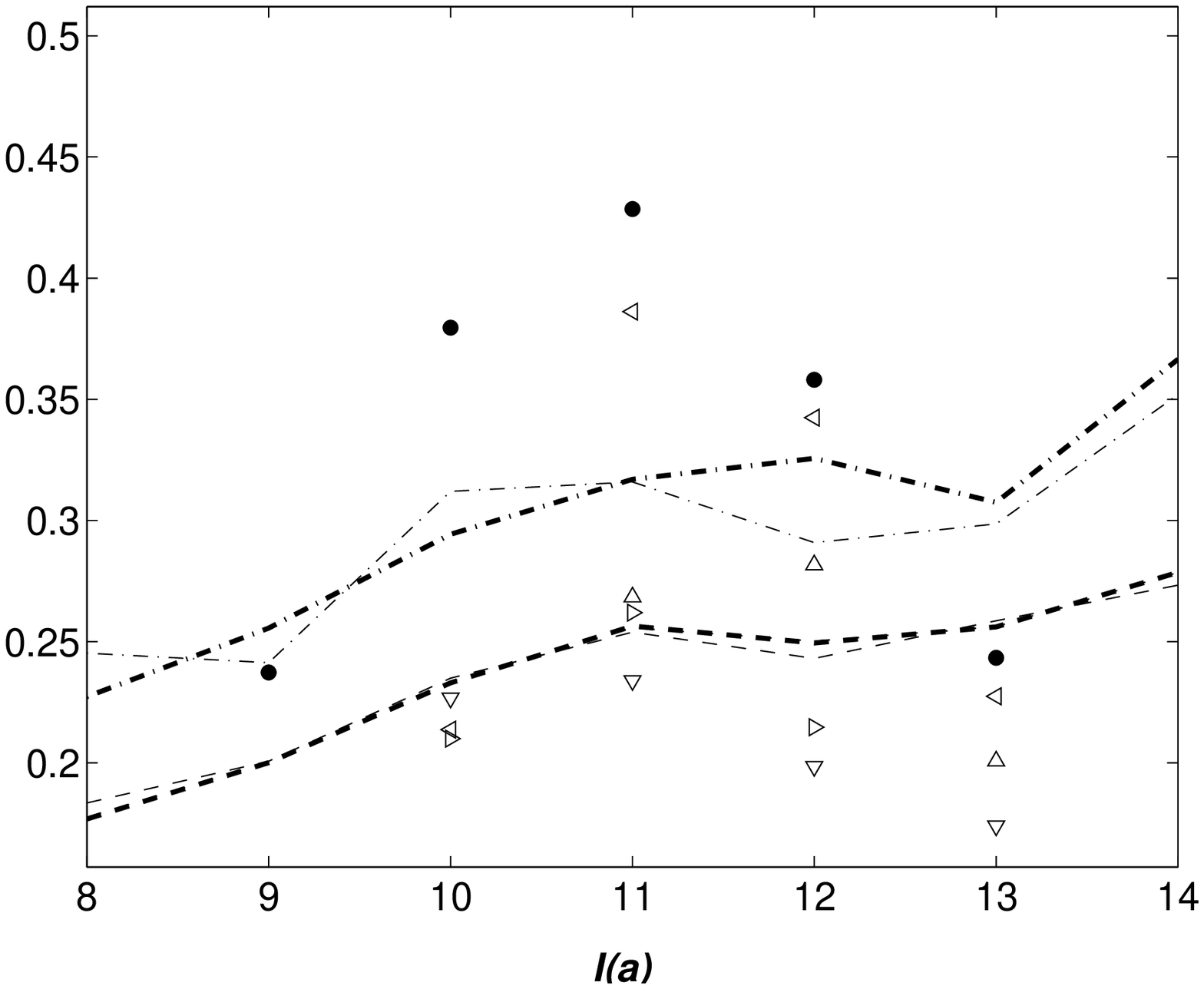}}
    \caption{Skewness values obtained from the analysis of each WMAP receiver.
    Panel ({\it a}) shows that all receivers exhibit a similar pattern with that of the combined map,
    indicating the non-Gaussianity is not caused by any particular receiver. However,
    the level of deviations are different for different bands; this is shown more clearly
    in ({\it b}), ({\it c}) and ({\it d}) which presents a magnified part of ({\it a}) separately from each band
    at scales of interest in this work.
    Since the noise and beams vary at some level among different receivers, we also carry out
    Monte Carlo simulations separately for each band. The heavy lines represent the confidence levels
    of each band, whereas the light lines show the confidence levels of the Q-V-W combined data as the
    same as those in ({\it a}) for a comparison. The W -band deviates less than the V -band, which deviates
    less than the Q -band, indicating the existence of some frequency-dependent signals
    that are most likely from the residual foreground contaminations. Note that Q -band is particularly suspected.}
\end{figure*}
\begin{deluxetable*}{rrrrrrrrr} \tablecolumns{9}
\tablecaption{Deviations and significance levels of spherical
wavelet coefficient statistics obtained from analyzing each-
receiver maps of the WMAP. Significant levels are calculated from
10,000 Gaussian simulations separately of each band. The number of
standard deviations the observation deviates from the mean is
given by $N_{\sigma}$; the corresponding significance level of the
detected non-Gaussianity is given by $\delta$.\label{tableeach}}
\tablehead{ \colhead{I(a)} & \multicolumn{2}{c}{10} & \colhead{} &
\multicolumn{2}{c}{11} & \colhead{}
& \multicolumn{2}{c}{12}\\
\cline{2-3} \cline{5-6} \cline{8-9} \\
\colhead{Receiver} & \colhead{$N_{\sigma}$} & \colhead{$\delta$} &
\colhead{}         & \colhead{$N_{\sigma}$} & \colhead{$\delta$} &
\colhead{}         & \colhead{$N_{\sigma}$} & \colhead{$\delta$}}
\startdata
Q1 & 3.4 &    99.9\% & & 3.6 & $>$99.9\% & & 2.9 & 99.5\% \\
Q2 & 3.7 & $>$99.9\% & & 4.0 & $>$99.9\% & & 3.0 & 99.6\% \\
V1 & 2.4 &    98.6\% & & 2.5 &    98.9\% & & 2.3 & 97.7\% \\
V2 & 3.4 &    99.9\% & & 3.3 &    99.9\% & & 2.6 & 98.8\% \\
W4 & 1.9 &    94.0\% & & 3.1 &    99.8\% & & 2.7 & 98.8\% \\
\enddata
\end{deluxetable*}

We now examine the latter possibility. Although skewness values
obtained from all receivers present very similar patterns, they
show some frequency dependence in Fig.~6. The deviation numbers of
each receiver and the corresponding significance levels from
simulations of each band are displayed in Table (\ref{tableeach}).
Although in one particular band, there are differences in the
relative amplitudes of each receivers, we may find a systematic
trend in frequency dependence among the three bands. Skewness of Q
-band deviates more than that of V -band, which also deviates more
than that of W -band. We have verified that these different levels
of deviations from Gaussian fluctuations are not caused by the
different measurement noises of these receivers at different
frequencies. The beams and noise have been calibrated out with
simulations of each individual band. The slight differences in
relative amplitudes from receivers of one particular band may be
due to subtle interplay between the instrument behavior and
foregrounds. If the central frequency of two receivers of the same
band differ from each other, then different levels of foregrounds
could remain in the different channels. Note that W -band would be
the band least contaminated with foreground synchrotron and
free-free emissions, and dust emissions has not reached its
maximum at the W-band frequency. We can take a rough quantitative
comparison of total foreground components among the three bands by
using Fig. 10a in Bennett {\it et al.} (2003b). This figure shows
that V -band and W -band would be cleaner than Q -band if
foreground residuals exist and that W -band would be the cleanest,
particularly if synchrotron-related foreground residuals exist.
Moreover, Fig.~7(b) shows that the skewness of the Tegmark cleaned
map exhibits less deviations from Gaussianity than the combined
WMAP map, perhaps due to more complete foreground removal in the
Tegmark map. Meanwhile, if the non-Gaussianity is totally due to
the intrinsic temperature fluctuations, there is no reason why the
deviations should be different at different frequencies and also
different between the {\it WMAP} team-recommended map and the
Tegmark map. In addition, noise properties cannot explain features
among different scale ranges. It might be true that due to
different noise properties the the deviations in the combined {\it
WMAP} map and the Tegmark cleaned map vary with each other on
small scales ($a_1\sim a_6$). But on medium scales $a_8\sim
a_{14}$, in which range the non-Gaussianity has been detected,
values of the {\it WMAP} map are systematically larger than those
of the Tegmark map. So if we believe that the Tegmark map is
cleaner than the {\it WMAP} map, the foreground residual
explanation is natural. We therefore conclude that the most likely
source of the detected non-Gaussianity is the residual foreground
signals in the map. As a cautious note, we cannot completely
exclude the intrinsic CMB signal in this work as the source of the
detected non-Gaussianity, since we have not determined the exact
nature of the possible foreground residual and quantified its
contribution to the non-Gaussianity systematically. Whether this
non-Gaussianity is due to foreground residuals or due to intrinsic
signals is still an open issue. Nevertheless, we have shown
several independent arguments for possible foreground residuals,
so it is highly likely that they are at least a promising
candidate for the source of the detected non-Gaussianity.

\begin{figure}
    \centering\subfigure[foregrounds and noise component]
    {\includegraphics[scale=0.39]{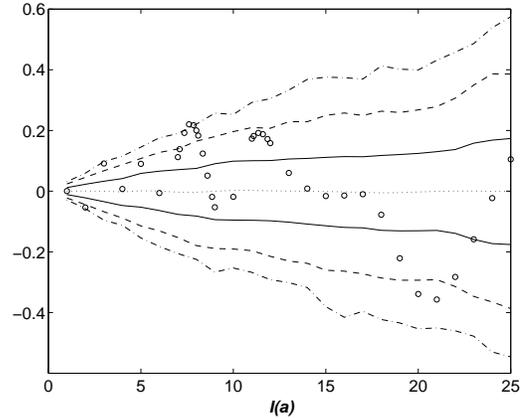}}
    \centering\subfigure[the combined, the Tegmark cleaned and a Gaussian CMB simulation with overestimated contamination]
    {\includegraphics[scale=0.39]{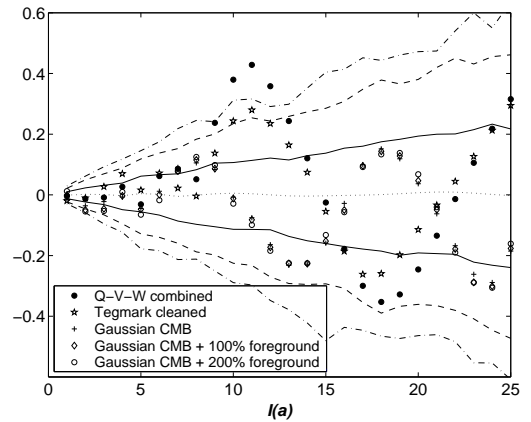}}
    \caption{Skewness values from the analysis of ({\it a}), foregrounds
    and noise component that is almost free of CMB emission, and
    ({\it b}), distinctly foreground-removed and overestimated {\it WMAP}-foreground-template contaminated maps.
    Panel ({\it a}) shows concordance with the Gaussian simulation and a different pattern
    from the CMB maps, indicating that the foreground component in the {\it WMAP} map does not contribute significantly to the detected
    non-Gaussianity. Panel ({\it b}) shows that the detected non-Gaussianity does not appear when adding to a
    Gaussian CMB simulation one and two times {\it WMAP}-foreground-template contaminations; deviations of the Tegmark map
    are less than those of the {\it WMAP} map, indicating some dependence
    on foreground removal technique, which could be caused by unresolved residual foreground.}
\end{figure}
%


\section{Conclusions and Discussions}

We have performed spherical wavelet analysis on the {\it WMAP}
first-year data and detected non-Gaussianity in wavelet space and
localized some non-Gaussian spots in real space. In wavelet space,
deviations from Gaussianity are found in skewness on scales from
$\sim11^{\circ}$ to $\sim14^{\circ}$, on the sky with a maximum
around $12^{\circ}$, and in kurtosis on scales smaller than
$\sim11^{\circ}$, with a maximum around $9^{\circ}$ toward
azimuthal orientation $\varphi_0 \sim 0^{\circ}$, using
anisotropic SMW. Several possible non-Gaussian spots are detected
in real space using isotropic SMHW from both the Q-V-W-combined
{\it WMAP} map and the Tegmark cleaned map. We have also
investigated the possible sources of this detected non-Gaussianity
in detail.

Systematic effects have been excluded, because (1) the skewness of
the maps including possible systematic instrumental features do
not show significant deviations from Gaussian fluctuations at the
scales where the non-Gaussianity is detected and (2) the skewness
is different at different frequencies and they all show very
different patterns from the CMB maps. To be stricter, there still
might be the possibility that some of the observations were due to
unknown systematic effects. Although we refer to different maps at
the same frequency as evidence of lack of systematics, this does
not exclude common-mode systematic artifacts. Foreground templates
adopted by the {\it WMAP} team also make no significant
contribution, because (1) the maps with significant contributions
from foregrounds and noise derived from the {\it WMAP} data do not
show significant deviations from Gaussian fluctuations and exhibit
very different patterns from CMB signals and (2) maps containing
overestimated {\it WMAP}-derived foreground template signals show
no deviation from Gaussianity. We find in several independent ways
that the level of non-Gaussianity is correlated positively with
the level of possible foreground contaminations due to incomplete
foreground removal. Among the three bands, Q -band seems to be the
most contaminated, probably due to residual synchrotron emission.
We conclude that the most likely source of the detected
non-Gaussianity is residual foreground signals in the map.

Although the {\it WMAP} team do not include the Q-band channels in
any power spectrum computations for $l<100$, which overlaps the
scales where the non-Gaussianity is detected, we want to remark
that the non-Gaussian features are also seen when excluding Q
-band. Individual band results primarily show that Q -band
contributes more to the non-Gaussianity than V and W bands, which
is reasonable for foreground residuals, since emissions are
dominant at some bands and minor at others. Therefore, it is still
premature to do more precise tests on the non-Gaussianity of the
intrinsic CMB fluctuations before we can identify the origin of
these foreground signals, understand their nature, and finally
remove them from the CMB maps completely. We will investigate the
origin of the detected non-Gaussian signals in our future work and
also look forward to the {\it WMAP} second-year data for
confirming these results and detecting these non-Gaussian signals
more precisely.




\noindent{\bf Acknowledgement: }
The authors kindly thank J. Laurent for very useful help on the
use of the YAWtb toolbox and M. Tegmark for guidance on the
weights of the Tegmark cleaned map. We are very grateful to the
anonymous referee whose extremely detailed and insightful comments
and suggestions allowed us to clarify several issues and improve
the readability of the paper.
We acknowledge the use of the LAMBDA. Support for LAMBDA is
provided by the NASA Office of Space Science.
This work has used the software package HEALPix (hierarchical,
equal area and iso-latitude pixelization of the sphere,
http://www.eso.org/science/healpix), developed by K.M. Gorski, E.
F. Hivon, B. D. Wandelt, J. Banday, F. K. Hansen and M.
Barthelmann.
This research has made use of the Astrophysical Integrated
Research Environment (AIRE) which is operated by the Center for
Astrophysics, Tsinghua University.
This study is supported in part by the Special Funds for Major State Basic Research
Projects, by the Directional Research Project of the Chinese Academy of Sciences and by
the National Natural Science Foundation of China.

\end{document}